\documentclass[aps,11pt]{revtex4}
\usepackage{dcolumn}
\usepackage{multirow}
\usepackage{array}
\usepackage{makecell}
\usepackage{graphicx}
\usepackage{amsmath}
\usepackage{amsfonts}
\usepackage{amssymb}
\usepackage{psfrag}
\usepackage{wrapfig}
\usepackage{subfigure}
\usepackage{makeidx}
\usepackage{bm}
\usepackage{color}
\usepackage{epsf}
\usepackage[hypertexnames=false]{hyperref}
\hypersetup{hidelinks}

\hypersetup{pdftitle={Quasinormal modes of type I perturbations of dilaton-Euler-Heisenberg black holes},pdfauthor={Yu-Xuan Wang, De-Cheng Zou, Sheng-Yuan Li, Yun Soo Myung, Ming Zhang}}
\usepackage{cases}
\usepackage{float}
\usepackage{placeins}
\usepackage{mathrsfs}

\makeatletter

\newcommand{\Rmnum}[1]{\uppercase\expandafter{\romannumeral #1}}
\makeatother

\newcommand{\dd}{\mathrm{d}}
\newcommand{\ii}{\mathrm{i}}
\newcommand{\rh}{r_{\rm h}}

\graphicspath{{figure/}}
\makeatletter
\renewcommand{\frontmatter@title@below}{\vskip12pt}
\renewcommand{\frontmatter@above@affiliation}{\vskip20pt}
\renewcommand{\frontmatter@abstractheading}{\begingroup\centering\normalsize\bfseries\abstractname\par\endgroup\vskip8pt}
\makeatother

\begin{document}

\title{Quasinormal modes of type I perturbations
for dilaton-Euler-Heisenberg black holes}

\author{Yu-Xuan Wang$^{1}$\footnote{wangyuxuan@jxnu.edu.cn},
De-Cheng Zou$^{1}$\footnote{corresponding author:dczou@jxnu.edu.cn},
Yun Soo Myung$^{2}$\footnote{ysmyung@inje.ac.kr},
Sheng-Yuan Li$^{3}$\footnote{shengyuanli77@outlook.com} and Ming Zhang$^{4}$\footnote{zhangming@xaau.edu.cn}}

\affiliation{$^{1}$School of Physics, Jiangxi Normal University, Nanchang 330022, China\\
$^{2}$Center for Quantum Spacetime, Sogang University, Seoul 04107, Republic of Korea\\
$^{3}$Center for Gravitation and Cosmology, College of Physical Science and Technology, Yangzhou University, Yangzhou 225009, China\\
$^{4}$Faculty of Science, Xihang University, Xi'an 710077, China}

\date{\today}

\begin{abstract}

\indent We study the type I quasinormal modes (QNMs) of magnetically charged dilaton--Euler--Heisenberg { (dEH)} black holes, where axial gravitational and polar electromagnetic perturbations are coupled.
Direct integration and Chebyshev pseudospectral methods give consistent QNM frequencies for the fundamental branches continued from Schwarzschild at $\ell=1,2,3$.
Varying $\alpha$ at fixed $\zeta=\alpha-\beta$ changes the spectrum on the same background.
The damping response depends on charge and multipole and can have opposite signs for the two coupled branches.
For $\zeta<0$, dipole damping decreases with charge toward horizon merger.
For $\zeta>0$, the Type I charge range extends beyond the complementary polarization's local angular-gradient bound.
All tracked modes remain damped in the nonextremal ranges, supporting the stability of these branches.

\end{abstract}
\maketitle

\section{Introduction}\label{sec:introduction}

General relativity has passed a wide range of experimental tests~\cite{Will:2014kxa}, and observations of binary black-hole mergers have extended these tests to the strong-field regime~\cite{LIGOScientific:2016aoc,LIGOScientific:2019fpa}. The ringdown signal allows us to examine the dynamics of the final black hole and possible additional degrees of freedom~\cite{Berti:2015itd}. Black-hole QNMs describe the characteristic damped oscillations of a perturbed black hole, with complex frequencies determined by the background, perturbation equations, and boundary conditions~\cite{Kokkotas:1999bd,Nollert:1999ji,Berti:2009kk,Konoplya:2011qq}. For an asymptotically flat black hole, these boundary conditions are ingoing waves at the event horizon and outgoing waves at infinity. With the time dependence $e^{-\ii\omega t}$, the real part of the frequency describes oscillation, {while a negative imaginary part $\omega_I<0$ gives exponential decay and a positive imaginary part $\omega_I>0$ gives exponential growth.} The measurement of several frequencies and damping times {constitutes} the basis of black-hole spectroscopy and tests of the no-hair prediction~\cite{Dreyer:2003bv,Berti:2005ys,Isi:2019aib}, although identifying these modes in a measured signal depends on their excitation and the ringdown model~\cite{Thrane:2017lqn,Bhagwat:2019dtm}.

Following the early studies of Schwarzschild perturbations~\cite{Regge:1957td,Zerilli:1970se}, black-hole QNMs have been investigated in scalar--tensor and higher-curvature theories, as well as in the presence of surrounding matter~\cite{Kobayashi:2012kh,Blazquez-Salcedo:2016enn,Franciolini:2018uyq,Cano:2021myl,Zhang:2024bov}. For charged black holes, gravitational and electromagnetic perturbations form a coupled system~\cite{Zerilli:1974ai,Moncrief:1974am}, with gravitational-led and electromagnetic-led branches distinguished by their neutral limits. The effects of scalar couplings on these modes have been studied for dilaton black holes~\cite{Ferrari:2000ep,Brito:2018hjh} and scalarized charged black holes in Einstein--Maxwell--scalar theory~\cite{Myung:2018jvi,Myung:2019oua}. At finite charge, both branches contain gravitational and electromagnetic perturbations, so their frequencies depend on the matter couplings as well as the black-hole geometry.

{ On the other hand,} the Euler--Heisenberg (EH) interaction has introduced nonlinear electromagnetic corrections from quantum electrodynamics~\cite{Heisenberg:1936nmg}, and black-hole solutions with these corrections have been studied~\cite{Yajima:2000kw,Ruffini:2013hia}. The EH interaction  allowed to modify wave propagation and the stability conditions~\cite{Obukhov:2002xa,Moreno:2002gg,Nomura:2020tpc}, and to break the parity isospectrality of charged black holes~\cite{Nomura:2021efi,Luo:2022EEHPerturbations}. {The magnetically charged dEH black hole solution~\cite{Bakopoulos:2024hah}  reduces to the Gibbons--Maeda or Garfinkle--Horowitz--Strominger solution when the EH correction disappears}~\cite{Gibbons:1987ps,Garfinkle:1990qj}. Previous work has considered external-field perturbations, axial gravitational modes, polar metric--dilaton perturbations, and the coupled Type II sector of these black holes~\cite{Zhang:2025xqt,Zou:2025MagneticAxial,Li:2026gqi,Li:TypeIICompanion}. Related studies have also examined QNMs and greybody factors in de Sitter backgrounds~\cite{Zhang:2026nog,Zhang:2026yfb}.

  {This paper focuses on studying the type I QNMs of magnetically charged dEH black holes. This is the remaining sector of the perturbation analysis of dEH black holes in the theory because it involves axial gravitational perturbations coupled to polar electromagnetic perturbations}~\cite{Nomura:2020tpc,Guo:2023CoupledQNM}. We derive the perturbation equations and calculate the frequencies by direct integration and Chebyshev pseudospectral methods~{\cite{Pani:2013pma,Jansen:2017oag,Lin:2016sch}}. We compare the two methods and examine the dependence of the oscillation frequencies and damping rates on magnetic charge and electromagnetic couplings.

The paper is organized as follows. Section~\ref{sec:background} presents the background solution and local electromagnetic conditions. {Section~\ref{sec:perturbations} is devoted to providing  the perturbation equations, and Sec.~\ref{sec:results} presents the QNM frequencies and numerical accuracy.} The conclusions are given in Sec.~\ref{sec:conclusion}. Furthermore, appendices~\ref{app:derivation} and \ref{sec:methods} describe the equation reduction and numerical methods, respectively.

\section{Dilaton--Euler--Heisenberg black holes}\label{sec:background}
\subsection{Field equations and black hole solution}\label{sec:field-equations}
In geometrized units $G=c=1$, the dilaton--Euler--Heisenberg theory considered in Ref.~\cite{Bakopoulos:2024hah} is described by the action
\begin{align}
S=\frac{1}{16\pi}\int\dd^4x\sqrt{-g}\Big\{
&R-2\nabla_\mu\phi\nabla^\mu\phi-e^{-2\phi}F_{\mu\nu}F^{\mu\nu}\nonumber\\
&-f(\phi)\left[2\alpha F^\mu{}_\nu F^\nu{}_\rho
F^\rho{}_\sigma F^\sigma{}_\mu
-\beta\left(F_{\mu\nu}F^{\mu\nu}\right)^2\right]\Big\},
\label{eq:action}
\end{align}
where $R$ is the Ricci scalar, $g=\det(g_{\mu\nu})$, $\phi$ is the dilaton, and $A_\mu$ is the electromagnetic potential. The field strength is defined as $F_{\mu\nu}=\partial_\mu A_\nu-\partial_\nu A_\mu$, and $\alpha$ and $\beta$ denote nonlinear electromagnetic coupling constants.

We define $\zeta=\alpha-\beta$ and introduce the dual field strength $\widetilde F^{\mu\nu}=\tfrac12\epsilon^{\mu\nu\rho\sigma}F_{\rho\sigma}$, where $\epsilon^{\mu\nu\rho\sigma}$ is the spacetime Levi--Civita tensor with the convention $\epsilon^{tr\theta\varphi}=-1/\sqrt{-g}$. Varying the action with respect to the metric, dilaton, and electromagnetic potential gives the Einstein, dilaton, and generalized Maxwell equations
\begin{align}
R_{\mu\nu}-\frac12R g_{\mu\nu}={}&2\partial_\mu\phi\partial_\nu\phi
-g_{\mu\nu}(\nabla\phi)^2
+2\left[e^{-2\phi}+2\zeta f(\phi)F_{\alpha\beta}F^{\alpha\beta}\right]F_{\mu\rho}F_\nu{}^\rho\nonumber\\
&-\frac12g_{\mu\nu}\left\{e^{-2\phi}F_{\alpha\beta}F^{\alpha\beta}
+f(\phi)\left[\zeta\left(F_{\alpha\beta}F^{\alpha\beta}\right)^2
-\frac{\alpha}{2}\left(F_{\alpha\beta}\widetilde F^{\alpha\beta}\right)^2\right]\right\},
\label{eq:Einsteinfield}\\
0={}&\Box\phi+\frac12e^{-2\phi}F_{\mu\nu}F^{\mu\nu}
-\frac14\frac{\dd f}{\dd\phi}\left[\zeta\left(F_{\mu\nu}F^{\mu\nu}\right)^2
+\frac{\alpha}{2}\left(F_{\mu\nu}\widetilde F^{\mu\nu}\right)^2\right],
\label{eq:dilatonfield}\\
0={}&\nabla_\mu\left\{\left[e^{-2\phi}+2\zeta f(\phi)F_{\alpha\beta}F^{\alpha\beta}\right]F^{\mu\nu}
+\alpha f(\phi)\left(F_{\alpha\beta}\widetilde F^{\alpha\beta}\right)\widetilde F^{\mu\nu}\right\}.
\label{eq:Maxwellfield}
\end{align}
Here $\Box=\nabla_\mu\nabla^\mu$. The static, spherically symmetric metric of the magnetic black hole is given by~\cite{Bakopoulos:2024hah}
\begin{equation}
\dd s^2=-A(r)\dd t^2+\frac{\dd r^2}{B(r)}+r^2\dd\Omega^2,
\label{eq:metric}
\end{equation}
where the metric functions are
\begin{align}
A(r)&=1-\frac{4M^2}{Q_m^2+\sqrt{Q_m^4+4M^2r^2}}
-\frac{2\zeta Q_m^4}{r^6},\label{eq:A}\\
B(r)&=\left(1+\frac{Q_m^4}{4M^2r^2}\right)A(r).\nonumber
\end{align}
Here $M$ and $Q_m$ are the mass and magnetic charge, respectively. The dilaton profile is
\begin{equation}
e^{2\phi(r)}=\frac{\sqrt{Q_m^4+4M^2r^2}+Q_m^2}
{\sqrt{Q_m^4+4M^2r^2}-Q_m^2}.\label{eq:Bphi}
\end{equation}
The dilaton coupling function is chosen as
\begin{equation}
f(\phi)=-3\cosh(2\phi)-2.
\label{eq:dilatoncoupling}
\end{equation}

At $\zeta=0$, the solution reduces to the magnetic GMGHS or GHS black hole~\cite{Gibbons:1987ps,Garfinkle:1990qj}, and at $Q_m=0$ to Schwarzschild. For nonzero charge, $\zeta>0$ gives one horizon. For $\zeta<0$, the solution can have two horizons, a degenerate horizon, or no horizon~\cite{Bakopoulos:2024hah,Zou:2025MagneticAxial}.

\subsection{Electromagnetic response and ghost-free regions}\label{sec:charge-domain}

Defining $X=F_{\mu\nu}F^{\mu\nu}$ and $Z=F_{\mu\nu}\widetilde F^{\mu\nu}$, the electromagnetic term in Eq.~\eqref{eq:action} is $-{\cal L}_{\rm EM}$, with
\begin{equation}
{\cal L}_{\rm EM}=e^{-2\phi}X+f(\phi)\left(\zeta X^2+\frac{\alpha}{2}Z^2\right).
\label{eq:EM-Lagrangian}
\end{equation}
On the magnetic background, $X=2Q_m^2/r^4$ and $Z=0$, so the background depends on the EH couplings only through $\zeta$, whereas the $\alpha Z^2$ term contributes at quadratic order. Thus varying $\alpha$ at fixed $M$, $Q_m$, and $\zeta$ changes the perturbations on the same geometry.

At fixed metric and dilaton, the quadratic electromagnetic term in a local orthonormal frame is
\begin{equation}
-{\cal L}^{(2)}_{\rm EM}=\frac{2e^{-2\phi}}{r^4}\Big[\Delta_1\left(\delta E_\perp^2-\delta B_\perp^2\right)+\Delta_2\delta E_\parallel^2-\Delta_3\delta B_\parallel^2\Big],
\label{eq:quadratic}
\end{equation}
where $\delta E$ and $\delta B$ denote electric and magnetic perturbations. The subscripts $\parallel$ and $\perp$ refer to the radial background magnetic field $Q_m/r^2$, with squared transverse amplitudes summed over the tangential directions. The response functions are
\begin{align}
\Delta_1&=r^4-2\zeta Q_m^2\left(3+4e^{2\phi}+3e^{4\phi}\right),\nonumber\\
\Delta_2&=r^4+2(\alpha-\zeta)Q_m^2\left(3+4e^{2\phi}+3e^{4\phi}\right),\nonumber\\
\Delta_3&=r^4-6\zeta Q_m^2\left(3+4e^{2\phi}+3e^{4\phi}\right).
\label{eq:D123}
\end{align}
The squared angular characteristic speeds are $c_{\rm I}^2=\Delta_1/\Delta_2$ and $c_{\rm II}^2=\Delta_3/\Delta_1$~\cite{Obukhov:2002xa,Nomura:2020tpc}. Positive type I kinetic and angular-gradient terms require
\begin{equation}
\Delta_1(r)>0,\qquad \Delta_2(r)>0,\qquad r\ge\rh.
\label{eq:typeI}
\end{equation}
Here $\rh$ is the outer horizon. Both polarizations additionally require $\Delta_3>0$ throughout the exterior. We take $\alpha\ge0$, so $\Delta_2\ge\Delta_1$ and the type I bound reduces to $\Delta_1>0$, independently of $\alpha$ at fixed $\zeta$. Full stability also requires an analysis of the coupled spectrum.

For $\zeta>0$ and nonzero charge, $\phi'<0$ implies $\Delta_1',\Delta_3'>0$, where primes denote radial derivatives. Their exterior minima therefore occur at the horizon. The type I and two-polarization charge limits follow from $A(\rh)=\Delta_1(\rh)=0$ and $A(\rh)=\Delta_3(\rh)=0$, respectively. At $\zeta/M^2=1$, they are
\begin{equation}
\frac{Q_{m,{\rm I}}}{M}=0.642549,\qquad
\frac{Q_{m,{\rm both}}}{M}=0.443105.
\label{eq:positive-charge-bounds}
\end{equation}
The horizon remains nonextremal at $Q_{m,{\rm I}}$. In the interval $Q_{m,{\rm both}}<Q_m<Q_{m,{\rm I}}$, type I satisfies its local conditions while $\Delta_3<0$ near the horizon; this interval is marked in the figures.

At $\zeta=0$, $\Delta_1=\Delta_3=r^4$ and $\Delta_2>0$, so the charge range is set by GHS regularity~\cite{Gibbons:1987ps,Garfinkle:1990qj}:
\begin{equation}
\rh=\sqrt{4M^2-2Q_m^2},\qquad |Q_m|<\sqrt{2}M.
\label{eq:GHSdomain}
\end{equation}
At the endpoint, the horizon area vanishes and the geometry becomes singular. For nonzero charge, pure Einstein--Maxwell--dilaton perturbations are recovered only for $\alpha=\beta=0$.

For $\zeta<0$, all three response functions are positive, and the upper charge bound is set by horizon merger, $A(\rh)=A'(\rh)=0$. At $\zeta/M^2=-1$, it is $Q_{m,{\rm ext}}/M=0.826114$~\cite{Bakopoulos:2024hah,Zou:2025MagneticAxial}.

\begin{figure}[htp]
\centering
\subfigure[\enspace $\zeta/M^2=1$]{
\includegraphics[width=0.32\textwidth]{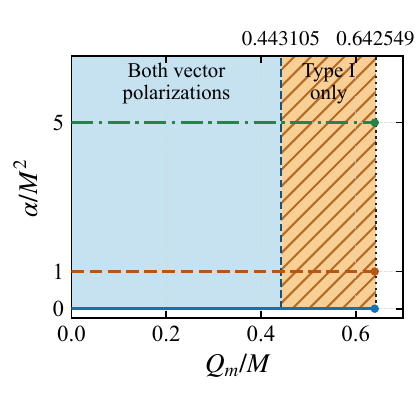}\label{fig:domain:a}}
\hfill
\subfigure[\enspace $\zeta/M^2=0$]{
\includegraphics[width=0.32\textwidth]{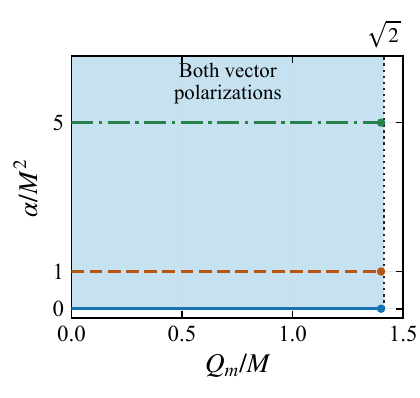}\label{fig:domain:b}}
\hfill
\subfigure[\enspace $\zeta/M^2=-1$]{
\includegraphics[width=0.32\textwidth]{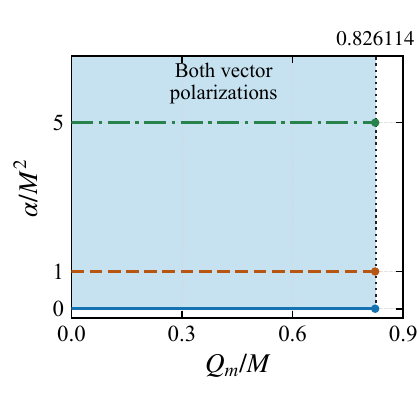}\label{fig:domain:c}}
\caption{Local electromagnetic domains for $M=1$ and $\alpha\ge0$. Blue shading denotes $\Delta_1,\Delta_2,\Delta_3>0$ throughout the exterior; orange hatching denotes the type I-only domain, with $\Delta_1,\Delta_2>0$ and $\Delta_3(\rh)<0$. Vertical lines mark excluded charge limits, and horizontal segments show the ranges at $\alpha/M^2=0,1,5$. The panels use different charge scales.}\label{fig:domain}
\end{figure}

Figure~\ref{fig:domain} summarizes these domains. The ranges end at $Q_m/M=0.64,1.4,0.824$ for $\zeta/M^2=1,0,-1$, respectively, below the corresponding limiting charges.

\FloatBarrier

\begin{samepage}
\section{Coupled perturbation equations}\label{sec:perturbations}
\subsection{Axial gravitational and polar electromagnetic perturbations}
The type I sector consists of coupled axial gravitational with polar electromagnetic perturbations~\cite{Nomura:2020tpc}. The metric and electromagnetic potential are expanded around dEH black hole background as 
\begin{equation}
g_{\mu\nu}=\bar g_{\mu\nu}+\epsilon h_{\mu\nu},\qquad
A_\mu=\bar A_\mu+\epsilon\delta A_\mu,
\label{eq:perturbedfields}
\end{equation}
\end{samepage}
where $\epsilon$ is a small parameter and overbars denote background fields, with $\bar A_\varphi=-Q_m\cos\theta$. In the type I sector, one finds that  $\delta X=0$, $\delta(\Box\bar\phi)=0$, and $\delta(Z^2)=2\overline Z\,\delta Z=0$. There is no  linearized dilaton mode as with $\delta\phi=0$. The variation $\delta Z$ can be nonzero and enters the linearized Maxwell equation, as shown in Appendix~\ref{app:derivation}. The indices $\ell$ and $m$ denote the angular multipole and azimuthal harmonic number. Spherical symmetry allows us to set $m=0$ without changing the frequencies. In the Regge--Wheeler gauge~\cite{Regge:1957td}, the axial metric perturbation is given by
\begin{equation}
h_{\mu\nu}=e^{-\ii\omega t}\sin\theta\,\partial_\theta Y_{\ell0}
\begin{pmatrix}
0&0&0&h_0(r)\\
0&0&0&h_1(r)\\
0&0&0&0\\
h_0(r)&h_1(r)&0&0
\end{pmatrix},
\label{eq:metricpert}
\end{equation}
while the polar electromagnetic perturbation, after removing its angular component by a gauge transformation, takes the form with two components $u_1$ and $u_2$
\begin{equation}
\delta A_\mu=e^{-\ii\omega t}Y_{\ell0}
\big(u_1(r),u_2(r),0,0\big).
\label{eq:EMpert}
\end{equation}
 Here $Y_{\ell0}$ is a scalar spherical harmonic, and $D_a$ is the covariant derivative on the unit two-sphere. The harmonics satisfy  eigenvalue equation: $D^aD_aY_{\ell0}=-\ell(\ell+1)Y_{\ell0}$. Four  functions $h_0,h_1,u_1,u_2$ are considered as radial amplitudes, and the tortoise coordinate is defined by $\dd r_*/\dd r=(AB)^{-1/2}$. After eliminating the constraints, the Einstein and Maxwell equations give one electromagnetic master equation for $\ell=1$ and two coupled equations for $\ell\ge2$. {We present the reduced equations and summarize the elimination steps in Appendix~\ref{app:derivation}.}

\begin{samepage}
\subsection{The \texorpdfstring{$\ell=1$}{ell=1} mode}
For $\ell=1$ mode,  it is clear that axial metric perturbations have no independent  degree of freedom~\cite{Moncrief:1974am,Martel:2005ir}. The Einstein and Maxwell constraints leave one electromagnetic master function, which for nonzero magnetic charge can be written as
\begin{equation}
\Psi_{\rm em}=r^4e^\phi\sqrt{\frac{B}{A\Delta_1}}
\left(h_0'+\ii\omega h_1-\frac{2h_0}{r}\right).
\label{eq:Gmaster}
\end{equation}
\end{samepage}
 {The relation between $\Psi_{\rm em}$ and the electromagnetic amplitudes, together with the reconstruction formulas, is given in Appendix~\ref{app:dipolereduction}.}
The resulting equation is
\begin{equation}
\frac{\dd^2\Psi_{\rm em}}{\dd r_*^2}
+\left[\omega^2-V_1(r)\right]\Psi_{\rm em}=0,
\label{eq:dipolemaster}
\end{equation}
where eliminating the metric constraints gives the effective potential
\begin{align}
V_1={}&AB\left[\phi''+\phi'^2-\frac{\Delta_1''}{2\Delta_1}
+\frac{3\Delta_1'^2}{4\Delta_1^2}
+\left(\frac{(AB)'}{2AB}+\frac4r-\frac{\Delta_1'}{\Delta_1}\right)\phi'\right]\nonumber\\
&+AB\left[\frac2{r^2}+\frac{(AB)'}{rAB}
-\left(\frac{(AB)'}{4AB}+\frac2r\right)\frac{\Delta_1'}{\Delta_1}\right]
+\frac{2A\Delta_1}{r^2}\left(\frac1{\Delta_2}
+\frac{2Q_m^2e^{-2\phi}}{r^6}\right).
\label{eq:dipolepotential}
\end{align}
The potential includes the metric response, although the dipole has only one radiative degree of freedom. In the neutral limit, it becomes
\begin{equation}
V_1\longrightarrow\frac{2A}{r^2}\qquad(Q_m\to0),
\label{eq:dipoleneutrallimit}
\end{equation}
which gives the Schwarzschild  dipole equation for $l=1$. The master variable in Eq.~\eqref{eq:Gmaster} and the reconstruction formulas in Appendix~\ref{app:dipolereduction} are defined for $Q_m\ne0$. At $Q_m=0$, we define the electromagnetic master variable directly from the Maxwell perturbations. The radial equation and potential have a smooth neutral limit.

\subsection{The \texorpdfstring{$\ell\ge2$}{ell>=2} modes}
For $\ell\ge2$, the metric and electromagnetic perturbations each have one  degree of freedom. Eliminating $h_0$ and $u_1$ leads to two coupled second-order equations for $h_1$ and $u_2$ as 
\begin{align}
0={}&h_1''+\left[\frac{3(AB)'}{2AB}-\frac2r\right]h_1'
-\frac{4Q_me^{-2\phi}\Delta_1}{r^6B}u_2\nonumber\\
&+\left[\frac{\omega^2}{AB}+\frac{(AB)''}{2AB}
-\frac{(AB)'}{rAB}+\frac{2-\ell(\ell+1)}{r^2B}
-\frac{4Q_m^2e^{-2\phi}\Delta_1}{r^8B}\right]h_1,
\label{eq:h1radial}\\
0={}&\frac{\dd}{\dd r}\left\{AB\left[u_2'
+\left(\frac{(AB)'}{2AB}-\frac4r-2\phi'
+\frac{\Delta_1'}{\Delta_1}\right)u_2\right]\right\}
+\omega^2u_2\nonumber\\
&-Q_m\frac{\dd}{\dd r}\left\{\frac{AB}{r^2}
\left(\frac6r+2\phi'-\frac{\Delta_1'}{\Delta_1}\right)h_1\right\}
-\frac{\ell(\ell+1)A\Delta_1}{r^4\Delta_2}
\left(Q_mh_1+r^2u_2\right).
\label{eq:u2radial}
\end{align}
The magnetic charge couples the metric and electromagnetic amplitudes in both equations. We choose  $\bm y=(Ah_1/r,Au_2)^{\rm T}$ as the numerical variables. The reconstruction formulas for $h_0$ and $u_1$ are given in Appendix~\ref{app:higherreduction}. At $Q_m=0$, these equations decouple, and two variables $Ah_1/r$ and $Au_2$ obey the Schwarzschild equations with their potentials
\begin{align}
V_{\rm GR}&=\left(1-\frac{2M}{r}\right)
\left[\frac{\ell(\ell+1)}{r^2}-\frac{6M}{r^3}\right],\nonumber\\
V_{\rm EM}&=\left(1-\frac{2M}{r}\right)\frac{\ell(\ell+1)}{r^2}.
\label{eq:Schwpotentials}
\end{align}
We call the branches continued from the Schwarzschild electromagnetic and gravitational modes as electromagnetic-led and gravitational-led, respectively. At finite charge, both branches contain metric and electromagnetic perturbations. These names refer to their Schwarzschild limits and do not imply a fixed ratio of the two amplitudes. We calculate the electromagnetic dipole and both coupled branches for $\ell=2,3$. Simultaneously changing the signs of $Q_m$, $u_1$, and $u_2$ leaves the radial equations unchanged, so we may restrict to $Q_m\ge0$.

\section{QNM frequencies}\label{sec:results}
We want to calculate the $\ell=1,2,3$ modes by making  use of direct integration (DI) and Chebyshev pseudospectral (PS) methods with $M=1$. The relevant numerical procedures are described in Appendix~\ref{sec:methods}. We take $\zeta/M^2=0,1,-1$ and $\alpha/M^2=0,1,5$  with $\beta=\alpha-\zeta$. At each $\zeta$, we compare different couplings at fixed mass and charge. The QNM boundary conditions are
described by 
\begin{align}
\Psi_{\rm em},\,Ah_1/r,\,Au_2&\sim e^{-\ii\omega r_*}\qquad(r\to\rh),\nonumber\\
\Psi_{\rm em},\,Ah_1/r,\,Au_2&\sim e^{+\ii\omega r_*}\qquad(r\to\infty).
\label{eq:QNMbc}
\end{align}
 We use $\Psi_{\rm em}$ for $\ell=1$ and  two components of $\bm y$ for $\ell\ge2$. With time dependence $e^{-\ii\omega t}$, we write the frequency as $\omega=\omega_R+\ii\omega_I$. The real part $\omega_R$ gives the oscillation frequency, and {$\omega_I<0$ describes decay with characteristic time $\tau=1/|\omega_I|$}. We consider  the fundamental branches continued from Schwarzschild one and label them  as $n=0$, where $n$ is the overtone index at zero charge. At finite charge, these branches need not be the least damped modes. Also,  we discuss the numerical accuracy in Sec.~\ref{sec:verification}.
{ The curves can be drawn by using the DI frequencies at the calculated charges without smoothing or extrapolation. We compute  the same QNM modes independently with the PS method.} The charge grid is refined near turning points and upper bounds. The quoted extrema are determined from the calculated frequencies.

\subsection{Electromagnetic dipole modes}\label{sec:dipole-results}

\begin{figure}[H]
\centering
\subfigure[\enspace $\zeta=0$]{
\includegraphics[width=0.485\textwidth]{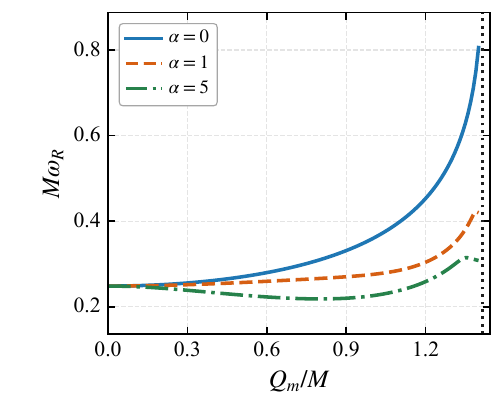}\label{fig:zerodipole:a}}
\hfill
\subfigure[\enspace $\zeta=0$]{
\includegraphics[width=0.485\textwidth]{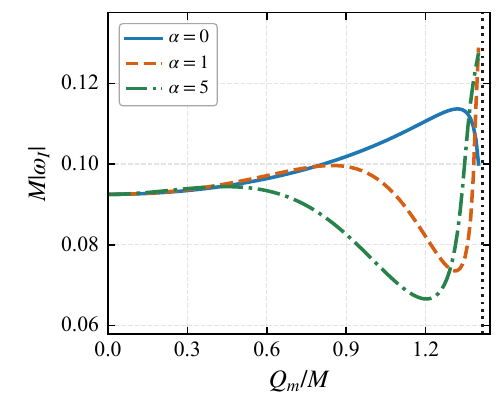}\label{fig:zerodipole:b}}
\caption{Electromagnetic dipole ($\ell=1$) QNM frequencies for $M=1$, $\zeta=0$, and $\alpha=\beta=0,1,5$. Two panels show the oscillation frequency and damping rate. The vertical dotted line marks the singular endpoint at $Q_m/M=\sqrt{2}$.}\label{fig:zerodipole}
\end{figure}

The dipole QNM frequencies for $\zeta=0$ are shown in Fig.~\ref{fig:zerodipole}. These theories share the magnetic GHS background, but their electromagnetic perturbation equations depend on $\alpha=\beta$. All curves start from the Schwarzschild frequency $M\omega_{\ell=1}=0.248263-0.092488\ii$ at $Q_m=0$~{\cite{Chandrasekhar:1975zza,Leaver:1986gd,Berti:2009kk}}. For $\alpha=0$, the real frequency increases throughout the calculated charge range. For $\alpha/M^2=5$, it first decreases below the Schwarzschild value and then increases. At $Q_m/M=1.4$, the frequencies are $M\omega=0.805833-0.100021\ii$ and $0.308043-0.127505\ii$ for $\alpha/M^2=0$ and 5, respectively. The real part decreases by $61.77\%$ on the same geometry.
The dipole potential depends on $\alpha$ only through $2A\Delta_1/(r^2\Delta_2)$ in Eq.~\eqref{eq:dipolepotential}, with $\zeta$ and charge held fixed. Increasing $\alpha$ raises $\Delta_2$ and lowers the potential throughout the allowed exterior. The real frequencies in Fig.~\ref{fig:zerodipole} also decrease. The damping rate has a different dependence on charge. At $Q_m/M=0.6$, increasing $\alpha/M^2$ from 0 to 5 reduces the real frequency by $20.33\%$ and the damping rate by $3.12\%$. For $\alpha/M^2=5$, the damping rate reaches a minimum of $0.066565$ at $Q_m/M=1.2$ and then increases sharply. At $Q_m/M=1.4$, it is $27.48\%$ larger than  $\alpha=0$ case, although the real frequency is lower than $\alpha=0$ case.

\begin{figure}[H]
\centering
\subfigure[\enspace $\zeta=1$]{
\includegraphics[width=0.485\textwidth]{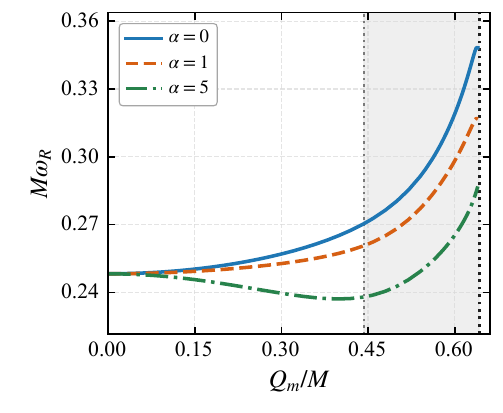}\label{fig:plusdipole:a}}
\hfill
\subfigure[\enspace $\zeta=1$]{
\includegraphics[width=0.485\textwidth]{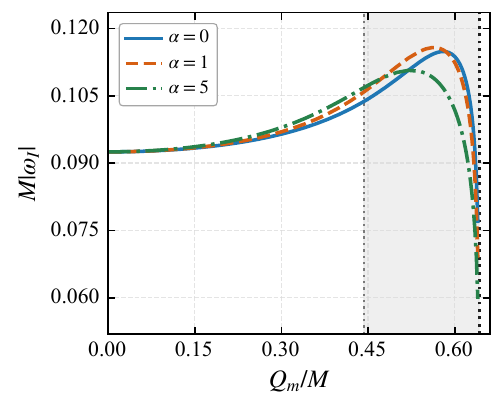}\label{fig:plusdipole:b}}
\caption{Electromagnetic dipole ($\ell=1$) QNM frequencies for $M=1$, $\zeta=1$, and $\alpha=0,1,5$. The panels show the oscillation frequency and damping rate. The gray band of $0.443105<Q_m/M<0.642549$ is suitable for the type I local condition only. Vertical dotted lines mark the two charge bounds.}\label{fig:plusdipole}
\end{figure}

In Fig.~\ref{fig:plusdipole}, the dipole damping rate peaks before the upper charge bound for $\zeta/M^2=1$. The maximum occurs at $Q_m/M=0.58$ for $\alpha=0$ and at $0.52$ for $\alpha/M^2=5$. Both  height and position of the maximum depend on $\alpha$. At $Q_m/M=0.4$, increasing $\alpha/M^2$ from 0 to 5 lowers the real frequency by $10.59\%$ and raises the damping rate by $2.83\%$. At $Q_m/M=0.6$, the damping rate decreases under the same coupling change. The damping curves therefore cross as the charge increases, whereas the ordering of the real-frequency curves remains unchanged.
The damping rate decreases near the upper charge bound, where $\Delta_1(\rh)$ approaches zero. At $Q_m/M=0.64$, the rates are $M|\omega_I|=0.077029$ and $0.059718$ for $\alpha/M^2=0$ and 5, respectively. The horizon remains nonextremal at this electromagnetic bound. In the gray band of $0.443105<Q_m/M<0.642549$, the type I conditions hold, but the complementary angular-gradient coefficient is negative near the horizon. The coupling effects at $Q_m/M=0.4$ occur below this band, where both polarizations satisfy the local conditions.

\begin{figure}[H]
\centering
\subfigure[\enspace $\zeta=-1$]{
\includegraphics[width=0.485\textwidth]{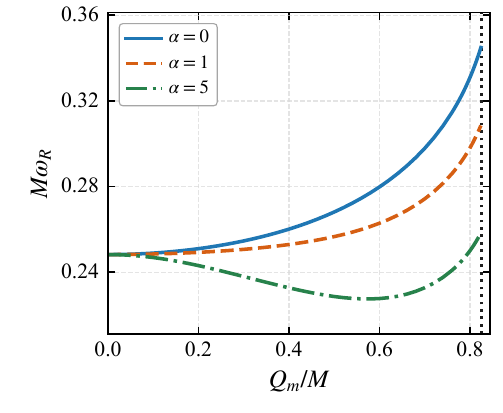}\label{fig:minusdipole:a}}
\hfill
\subfigure[\enspace $\zeta=-1$]{
\includegraphics[width=0.485\textwidth]{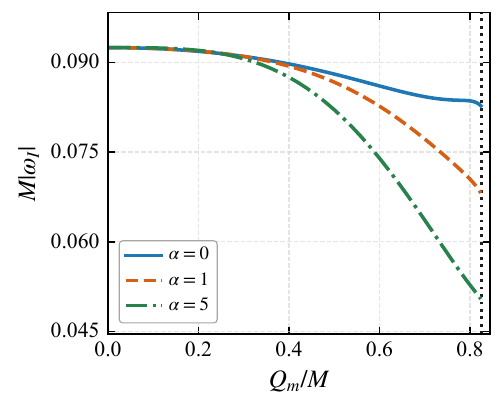}\label{fig:minusdipole:b}}
\caption{Electromagnetic dipole ($\ell=1$) QNM frequencies for $M=1$, $\zeta=-1$, and $\alpha=0,1,5$. The panels show the oscillation frequency and damping rate. The vertical dotted line represents  the horizon-merger endpoint at $Q_m/M=0.826114$.}\label{fig:minusdipole}
\end{figure}

For $\zeta/M^2=-1$, the dipole damping rate decreases with increasing charge for each coupling, as shown in Fig.~\ref{fig:minusdipole}. Increasing $\alpha$ lowers the damping rate further at moderate and large charge. For $\alpha/M^2=5$, the real frequency has a shallow minimum near $Q_m/M=0.58$ and then increases. The damping rate continues to decrease, so the decaying time grows even as the oscillation frequency rises.
Increasing $\alpha/M^2$ from 0 to 5 lowers the real frequency by $18.63\%$ and the damping rate by $14.06\%$ at $Q_m/M=0.6$. At $Q_m/M=0.824$, the corresponding frequencies are $M\omega=0.344911-0.082746\ii$ and $0.258387-0.050433\ii$. The damping rate decreases by $39.05\%$, while the real part decreases by $25.09\%$. All electromagnetic coefficients remain positive on this branch, whose upper charge bound is set by horizon merger. The last calculated point lies below this bound. Fig.~\ref{fig:complexcomparison:a} also shows the different relations between oscillation frequency and damping rate on the three backgrounds.

\FloatBarrier

\subsection{Coupled quadrupole and octupole modes}\label{sec:quadrupole-results}

\begin{figure}[H]
\centering
\subfigure[{\enspace $\zeta=0$, $\omega_R$, G-led}]{
\includegraphics[width=0.42\textwidth]{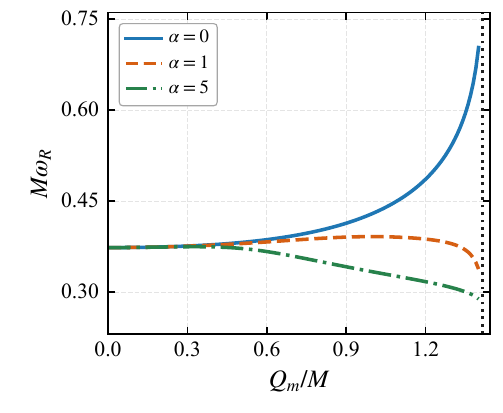}\label{fig:zeroquadrupole:a}}
\hfill
\subfigure[{\enspace $\zeta=0$, $|\omega_I|$, G-led}]{
\includegraphics[width=0.42\textwidth]{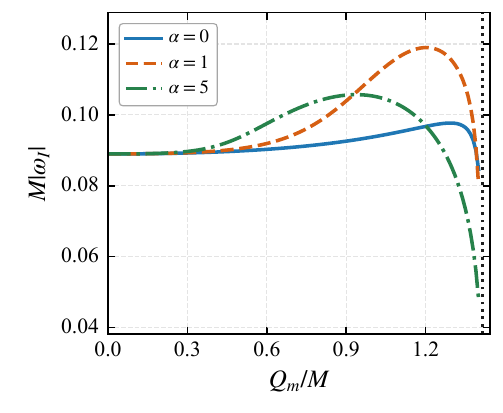}\label{fig:zeroquadrupole:b}}
\par\vspace{3pt}
\subfigure[{\enspace $\zeta=0$, $\omega_R$, EM-led}]{
\includegraphics[width=0.42\textwidth]{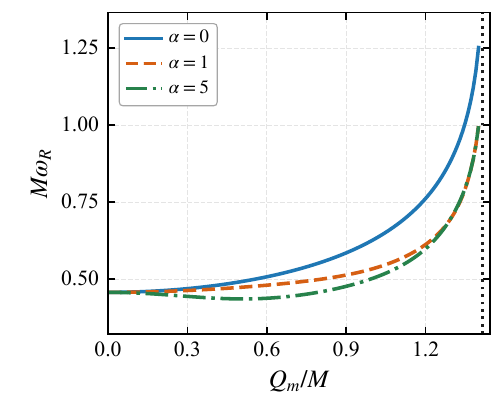}\label{fig:zeroquadrupole:c}}
\hfill
\subfigure[{\enspace $\zeta=0$, $|\omega_I|$, EM-led}]{
\includegraphics[width=0.42\textwidth]{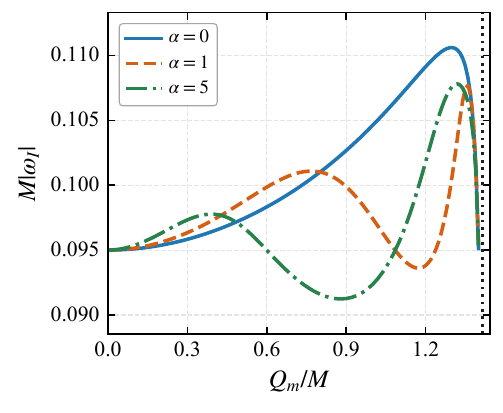}\label{fig:zeroquadrupole:d}}
\caption{Coupled quadrupole ($\ell=2$) QNM frequencies for $M=1$, $\zeta=0$, and $\alpha=\beta=0,1,5$. The upper and lower rows show the G-led and EM-led branches, respectively. The columns give the oscillation frequency and damping rate. The vertical dotted line marks $Q_m/M=\sqrt{2}$.}\label{fig:zeroquadrupole}
\end{figure}

The two quadrupole branches respond differently to the coupling on the GHS background, as shown in Fig.~\ref{fig:zeroquadrupole}. Their Schwarzschild frequencies are known to be $M\omega_{\ell=2}^{\rm grav}=0.373672-0.088962\ii$ and $M\omega_{\ell=2}^{\rm em}=0.457596-0.095004\ii$~{\cite{Chandrasekhar:1975zza,Leaver:1986gd,Berti:2009kk}}. The superscripts denote the gravitational (G)-led and electromagnetic (EM)-led branches. At $Q_m/M=0.6$, increasing $\alpha=\beta$ from 0 to $5M^2$ lowers their real frequencies by $5.07\%$ and $13.71\%$, respectively. At $Q_m/M=1.4$, the G-led frequency changes from $0.703216-0.084648\ii$ to $0.289074-0.048037\ii$. The EM-led frequency changes from $1.251989-0.095120\ii$ to $0.996868-0.095367\ii$. The corresponding real-frequency reductions are $58.89\%$ and $20.38\%$. The G-led branch is therefore more sensitive to $\alpha$ at large charge, although it is less sensitive at moderate charge.
{ We emphasize that Eq.~\eqref{eq:h1radial} has no explicit $\alpha$ dependence at fixed background. However, the last term in Eq.~\eqref{eq:u2radial} makes $\alpha$-coupling to electromagnetic equation through $\Delta_2$.}  The G-led frequency therefore changes when $\alpha$ is varied. At finite charge, this frequency belongs to the coupled system, even though the branch starts from a gravitational mode at $Q_m=0$.
The same coupling change at $Q_m/M=0.6$ increases the G-led damping rate by $8.06\%$ and decreases the EM-led rate by $3.40\%$. For $\alpha/M^2=0,1,5$, the G-led damping rates reach maxima of $M|\omega_I|=0.097690,0.118992,0.105706$ at $Q_m/M=1.29,1.2,0.94$, respectively. The maximum shifts to lower charge as $\alpha$ increases, and its height is largest for $\alpha/M^2=1$. At $Q_m/M=1.4$, increasing $\alpha/M^2$ from 0 to 5 reduces the G-led damping rate by $43.25\%$. The EM-led rates differ by only $0.26\%$, although their real frequencies differ appreciably.

\begin{figure}[H]
\centering
\subfigure[{\enspace $\omega_R$, G-led}]{
\includegraphics[width=0.42\textwidth]{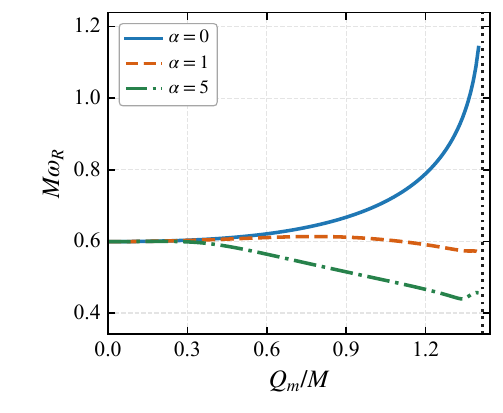}\label{fig:zerooctupole:a}}
\hfill
\subfigure[{\enspace $|\omega_I|$, G-led}]{
\includegraphics[width=0.42\textwidth]{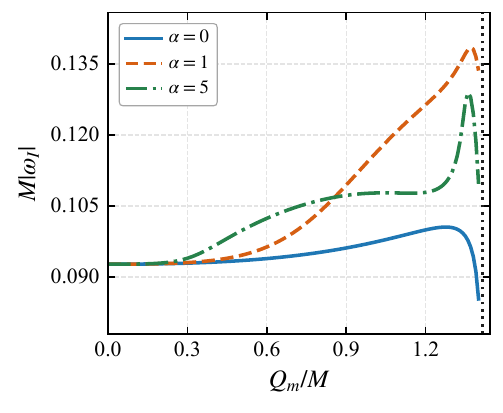}\label{fig:zerooctupole:b}}
\par\vspace{3pt}
\subfigure[{\enspace $\omega_R$, EM-led}]{
\includegraphics[width=0.42\textwidth]{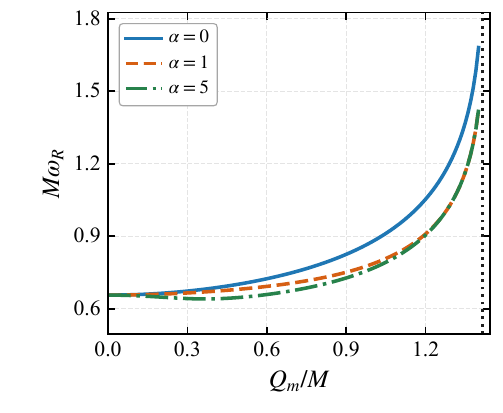}\label{fig:zerooctupole:c}}
\hfill
\subfigure[{\enspace $|\omega_I|$, EM-led}]{
\includegraphics[width=0.42\textwidth]{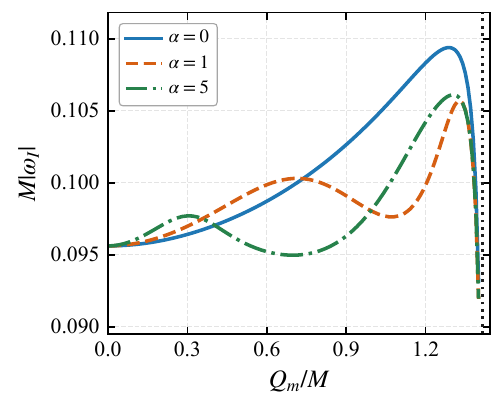}\label{fig:zerooctupole:d}}
\par\vspace{3pt}
\caption{QNM frequencies of the coupled $\ell=3$ modes for $M=1$, $\zeta=0$, and $\alpha=0,1,5$. The upper and lower rows show the G-led and EM-led branches, respectively. The columns give the oscillation frequency and damping rate. The vertical dotted line marks $Q_m/M=\sqrt2$.}\label{fig:zerooctupole}
\end{figure}

As shown in Fig.~\ref{fig:zerooctupole}, similar real-frequency shifts occur at moderate charge for the two $\ell=3$ branches on the GHS background. Their Schwarzschild frequencies are $M\omega_{\ell=3}^{\rm grav}=0.599443-0.092703\ii$ and $M\omega_{\ell=3}^{\rm em}=0.656899-0.095616\ii$~{\cite{Chandrasekhar:1975zza,Leaver:1986gd,Berti:2009kk}}. At $Q_m/M=0.6$, increasing $\alpha/M^2$ from 0 to 5 reduces the G-led and EM-led real frequencies by $9.15\%$ and $9.01\%$, respectively. The two reductions are nearly equal, whereas the quadrupole reductions in Fig.~\ref{fig:zeroquadrupole} differ. The damping shifts have opposite signs, with an increase of $9.05\%$ for the G-led branch and a decrease of $3.55\%$ for the EM-led branch.
For $\alpha/M^2=5$ and $Q_m/M=1.4$, the octupole frequencies are $M\omega^{\rm grav}_{\ell=3}=0.456471-0.109674\ii$ and $M\omega^{\rm em}_{\ell=3}=1.424042-0.091955\ii$. Both real frequencies are higher than those of the corresponding quadrupoles at the same $\alpha$, $\zeta$, and charge. The G-led octupole has a larger damping rate than the quadrupole, while the EM-led octupole has a smaller rate. At this charge, increasing $\alpha/M^2$ from 0 to 5 raises the G-led octupole damping rate by $28.47\%$. This is opposite to the $43.25\%$ decrease for $\ell=2$. The damping response thus depends on the multipole even for the same background and branch.

\begin{figure}[H]
\centering
\subfigure[{\enspace $\zeta=1$, $\omega_R$, G-led}]{
\includegraphics[width=0.42\textwidth]{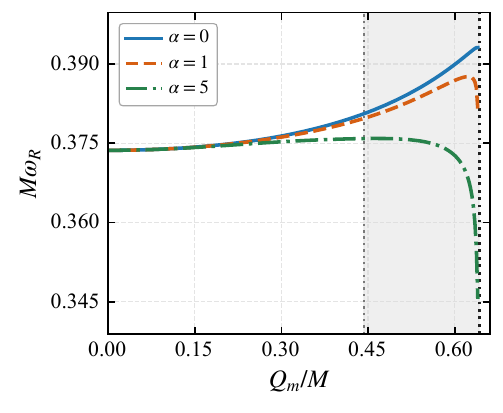}\label{fig:plusquadrupole:a}}
\hfill
\subfigure[{\enspace $\zeta=1$, $|\omega_I|$, G-led}]{
\includegraphics[width=0.42\textwidth]{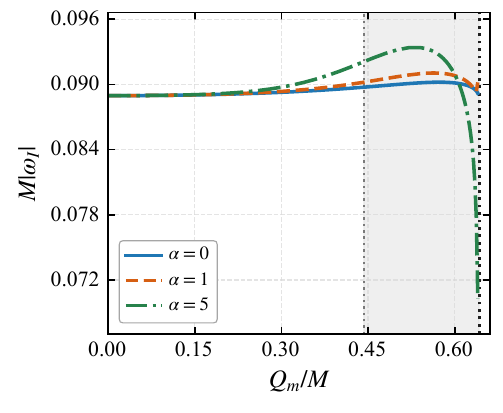}\label{fig:plusquadrupole:b}}
\par\vspace{3pt}
\subfigure[{\enspace $\zeta=1$, $\omega_R$, EM-led}]{
\includegraphics[width=0.42\textwidth]{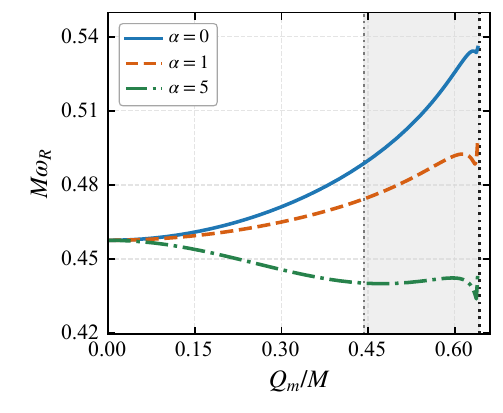}\label{fig:plusquadrupole:c}}
\hfill
\subfigure[{\enspace $\zeta=1$, $|\omega_I|$, EM-led}]{
\includegraphics[width=0.42\textwidth]{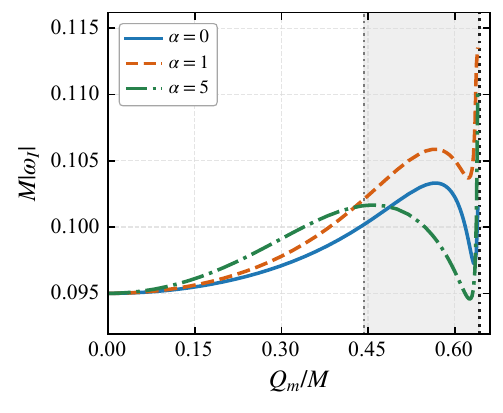}\label{fig:plusquadrupole:d}}
\caption{Coupled quadrupole ($\ell=2$) QNM frequencies for $M=1$, $\zeta=1$, and $\alpha=0,1,5$. The upper and lower rows show the G-led and EM-led branches. The columns give the oscillation frequency and damping rate. The gray band denotes the type I-only range defined in Fig.~\ref{fig:plusdipole}.}\label{fig:plusquadrupole}
\end{figure}

The G-led quadrupole real frequencies for $\zeta>0$ remain close at moderate charge, while the EM-led frequencies are more separated, as shown in  Fig.~\ref{fig:plusquadrupole}. At $Q_m/M=0.4$, increasing $\alpha/M^2$ from 0 to 5 lowers their real frequencies by $0.85\%$ and $8.54\%$, respectively. Their damping rates increase by $1.81\%$ and $2.14\%$. At this charge, the two branches differ mainly in the coupling dependence of the real frequency.
Near the type I charge bound, the G-led frequency changes more strongly with $\alpha$. At $Q_m/M=0.64$, it changes from $0.393114-0.089181\ii$ to $0.345556-0.070248\ii$ when $\alpha/M^2$ increases from 0 to 5. The real part and damping rate decrease by $12.10\%$ and $21.23\%$, respectively. For the EM-led branch, the same coupling change lowers the damping rate at $Q_m/M=0.6$. At $Q_m/M=0.64$, it instead raises the rate from $0.101458$ to $0.110421$. The damping dependence on $\alpha$ is non-monotonic at $Q_m/M=0.6$, where $\alpha/M^2=1$ gives a larger rate than either 0 or 5. At $Q_m/M=0.64$, increasing $\alpha/M^2$ from 0 to 5 lengthens the G-led decay time and shortens the EM-led one. These damping shifts follow from the coupled equations and cannot be inferred from the decrease of $\Delta_1/\Delta_2$ alone.

\begin{figure}[H]
\centering
\subfigure[{\enspace $\omega_R$, G-led}]{
\includegraphics[width=0.42\textwidth]{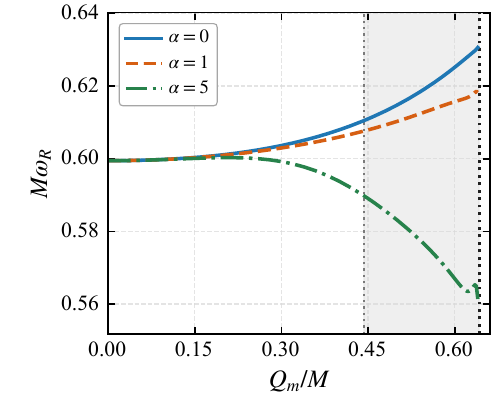}\label{fig:plusoctupole:a}}
\hfill
\subfigure[{\enspace $|\omega_I|$, G-led}]{
\includegraphics[width=0.42\textwidth]{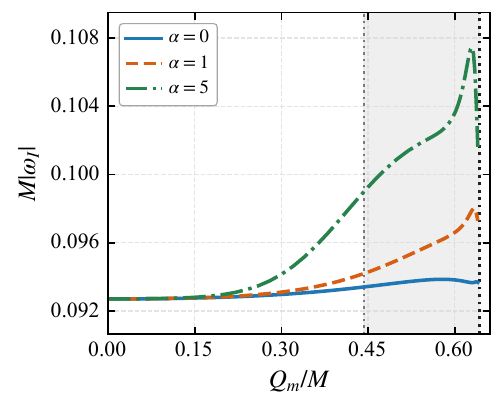}\label{fig:plusoctupole:b}}
\par\vspace{3pt}
\subfigure[{\enspace $\omega_R$, EM-led}]{
\includegraphics[width=0.42\textwidth]{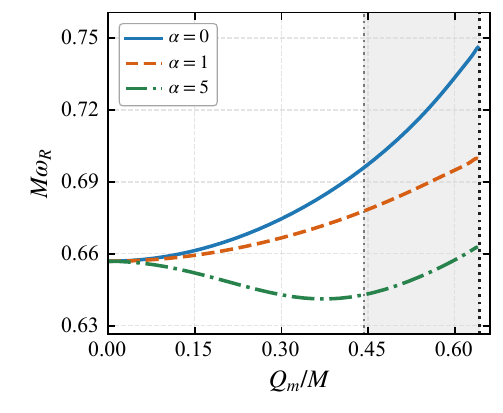}\label{fig:plusoctupole:c}}
\hfill
\subfigure[{\enspace $|\omega_I|$, EM-led}]{
\includegraphics[width=0.42\textwidth]{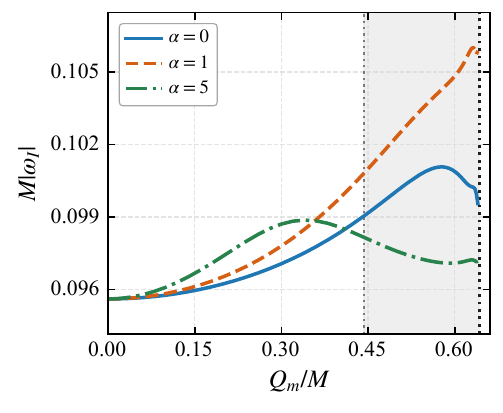}\label{fig:plusoctupole:d}}
\par\vspace{3pt}
\caption{QNM frequencies of the coupled $\ell=3$ modes for $M=1$, $\zeta=1$, and $\alpha=0,1,5$. The upper and lower rows show the G-led and EM-led branches, respectively. The columns give the oscillation frequency and damping rate. The gray band denotes the Type I-only charge range.}\label{fig:plusoctupole}
\end{figure}

Figure~\ref{fig:plusoctupole} extends the comparison for $\zeta>0$ to $\ell=3$. At $Q_m/M=0.4$, increasing $\alpha/M^2$ from 0 to 5 reduces the G-led and EM-led real frequencies by $2.32\%$ and $6.84\%$, respectively. The G-led reduction is larger than for $\ell=2$, whereas the EM-led reduction is smaller. The damping rates increase by $4.36\%$ and $0.25\%$. The EM-led damping rate changes little at this charge, although its real frequency decreases.
The G-led real frequency at $\alpha/M^2=5$ falls below the Schwarzschild value as the charge increases. The EM-led real frequency first reaches a shallow minimum and then rises. At $Q_m/M=0.64$, the frequencies are $M\omega^{\rm grav}_{\ell=3}=0.561419-0.101622\ii$ and $M\omega^{\rm em}_{\ell=3}=0.662909-0.097186\ii$. Their damping rates are $8.48\%$ larger and $2.39\%$ smaller, respectively, than for $\alpha=0$ at the same charge. Both changes have the opposite signs to those for $\ell=2$. The $\ell=3$ damping curves also turn near the upper charge bound, but the G-led branch has no strong damping suppression as for $\ell=2$.

\begin{figure}[H]
\centering
\subfigure[{\enspace $\zeta=-1$, $\omega_R$, G-led}]{
\includegraphics[width=0.42\textwidth]{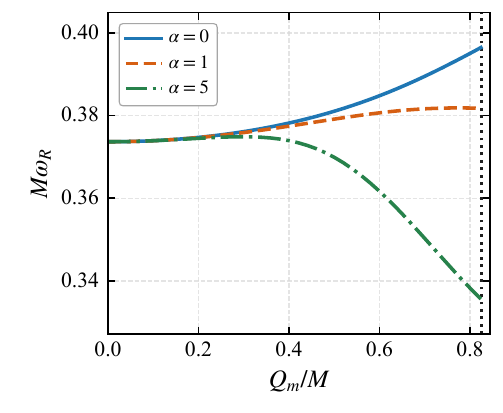}\label{fig:minusquadrupole:a}}
\hfill
\subfigure[{\enspace $\zeta=-1$, $|\omega_I|$, G-led}]{
\includegraphics[width=0.42\textwidth]{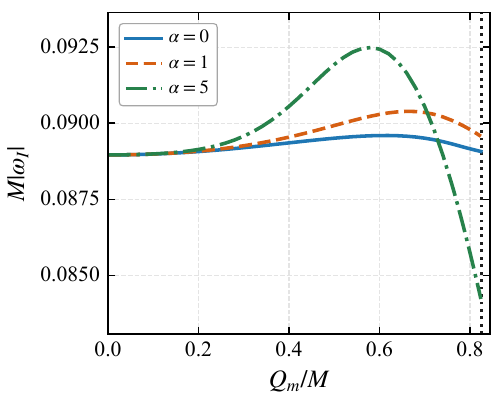}\label{fig:minusquadrupole:b}}
\par\vspace{3pt}
\subfigure[{\enspace $\zeta=-1$, $\omega_R$, EM-led}]{
\includegraphics[width=0.42\textwidth]{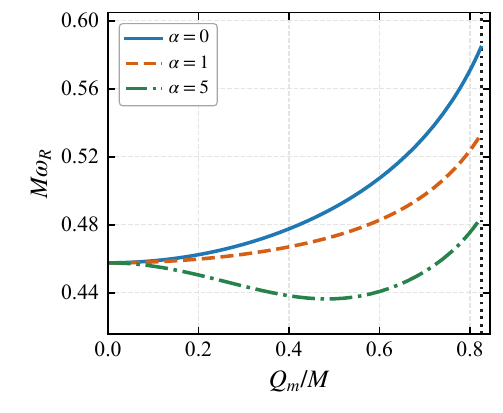}\label{fig:minusquadrupole:c}}
\hfill
\subfigure[{\enspace $\zeta=-1$, $|\omega_I|$, EM-led}]{
\includegraphics[width=0.42\textwidth]{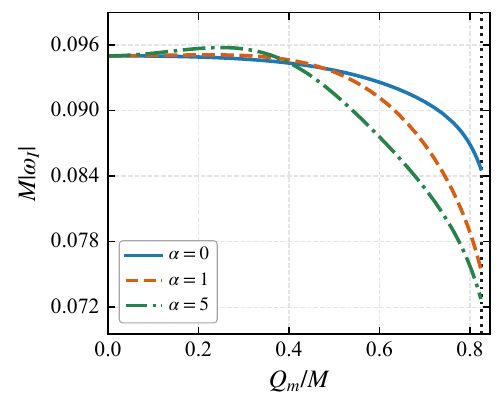}\label{fig:minusquadrupole:d}}
\caption{Coupled quadrupole ($\ell=2$) QNM frequencies for $M=1$, $\zeta=-1$, and $\alpha=0,1,5$. The upper and lower rows show the G-led and EM-led branches. The columns give the oscillation frequency and damping rate. The vertical dotted line marks $Q_m/M=0.826114$.}\label{fig:minusquadrupole}
\end{figure}

The quadrupole branches for $\zeta<0$ exhibit different turning points, as shown in  Fig.~\ref{fig:minusquadrupole}. For $\alpha/M^2=5$, the G-led real frequency increases slightly at small charge and then decreases. The EM-led real frequency reaches a minimum near $Q_m/M=0.48$ and then increases. The G-led damping rate reaches a maximum near $Q_m/M=0.58$. The real frequency and damping rate therefore turn at different charges.
At $Q_m/M=0.6$, increasing $\alpha/M^2$ from 0 to 5 lowers the G-led and EM-led real frequencies by $6.01\%$ and $13.14\%$, respectively. Their damping rates change by $+3.17\%$ and $-5.41\%$. At $Q_m/M=0.824$, the G-led frequency changes from $0.396461-0.089076\ii$ to $0.335645-0.084231\ii$. The EM-led frequency changes from $0.584217-0.084660\ii$ to $0.484161-0.072751\ii$. Both damping rates decrease at this larger charge. The G-led damping shift thus changes sign between the two charges, while the EM-led shift remains negative. Both points satisfy the local conditions for the two electromagnetic polarizations and lie below the horizon-merger bound.

\begin{figure}[H]
\centering
\subfigure[{\enspace $\omega_R$, G-led}]{
\includegraphics[width=0.42\textwidth]{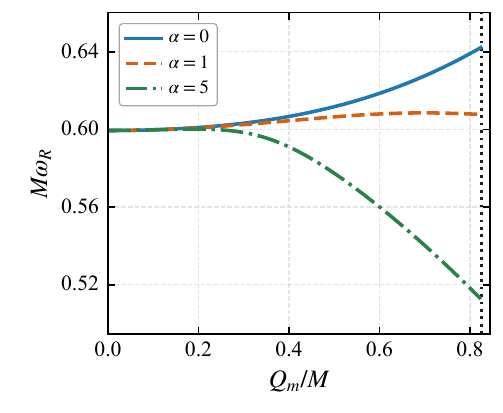}\label{fig:minusoctupole:a}}
\hfill
\subfigure[{\enspace $|\omega_I|$, G-led}]{
\includegraphics[width=0.42\textwidth]{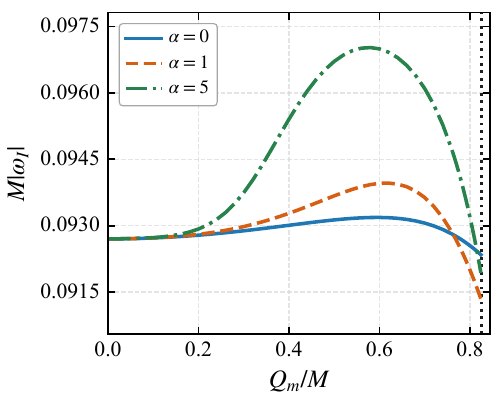}\label{fig:minusoctupole:b}}
\par\vspace{3pt}
\subfigure[{\enspace $\omega_R$, EM-led}]{
\includegraphics[width=0.42\textwidth]{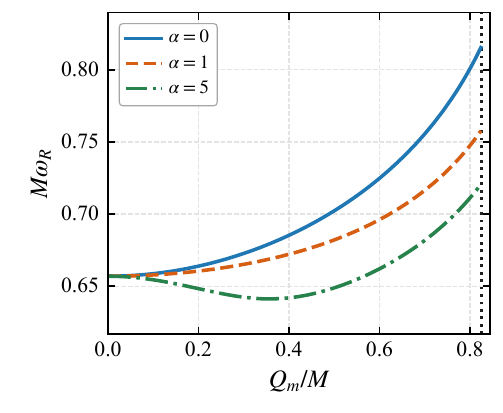}\label{fig:minusoctupole:c}}
\hfill
\subfigure[{\enspace $|\omega_I|$, EM-led}]{
\includegraphics[width=0.42\textwidth]{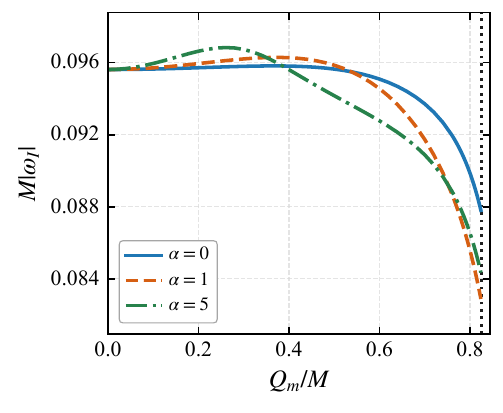}\label{fig:minusoctupole:d}}
\par\vspace{3pt}
\caption{QNM frequencies of the coupled $\ell=3$ modes for $M=1$, $\zeta=-1$, and $\alpha=0,1,5$. The upper and lower rows show the G-led and EM-led branches, respectively. The columns give the oscillation frequency and damping rate. The vertical dotted line marks $Q_m/M=0.826114$.}\label{fig:minusoctupole}
\end{figure}

Increasing the multipole to $\ell=3$ changes the relative coupling sensitivity of the two branches for $\zeta<0$, as shown in Fig.~\ref{fig:minusoctupole}. At $Q_m/M=0.6$, increasing $\alpha/M^2$ from 0 to 5 lowers the G-led and EM-led real frequencies by $9.46\%$ and $8.66\%$, respectively. The G-led shift is slightly larger, reversing the ordering for $\ell=2$ at the same charge. The damping rates change by $+4.09\%$ and $-2.44\%$, so their shifts retain the opposite signs found for the quadrupoles.
Close to horizon merger at $Q_m/M=0.824$, increasing $\alpha/M^2$ from 0 to 5 lowers the G-led octupole real frequency by $20.17\%$. Its damping rate decreases by only $0.44\%$. For $\alpha/M^2=5$, the frequencies are $M\omega^{\rm grav}_{\ell=3}=0.512653-0.091934\ii$ and $M\omega^{\rm em}_{\ell=3}=0.720572-0.084344\ii$. Both octupoles have higher real frequencies and larger damping rates than the corresponding quadrupoles. At this charge, both $\alpha/M^2=1$ octupoles have slightly smaller damping rates than their $\alpha/M^2=5$ counterparts. Increasing the coupling can therefore shorten the decay time while lowering the real frequency.

\begin{figure}[H]
\centering
\subfigure[\enspace $\ell=1$]{
\includegraphics[width=.32\textwidth]{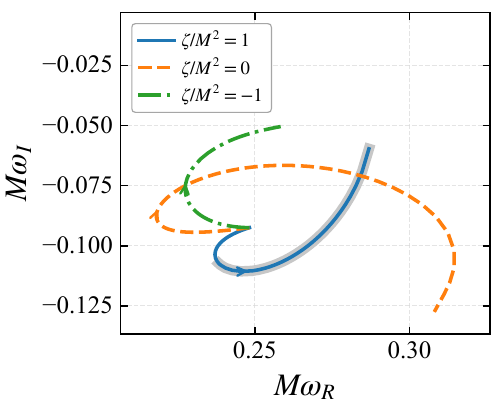}\label{fig:complexcomparison:a}}
\hfill
\subfigure[\enspace $\ell=2$, G-led]{
\includegraphics[width=.32\textwidth]{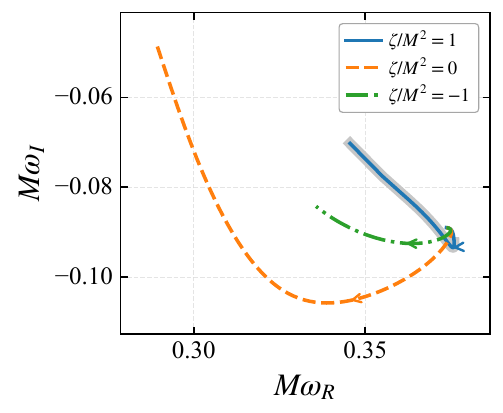}\label{fig:complexcomparison:b}}
\hfill
\subfigure[\enspace $\ell=2$, EM-led]{
\includegraphics[width=.32\textwidth]{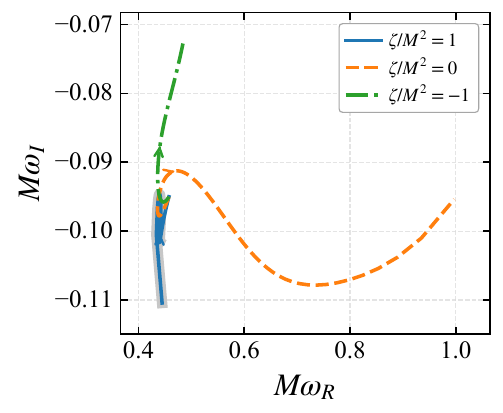}\label{fig:complexcomparison:c}}
\caption{Complex-frequency trajectories for $M=1$, $\alpha/M^2=5$, and $\zeta/M^2=1,0,-1$. Solid, dashed, and dash-dotted curves denote $\zeta/M^2=1,0,-1$, respectively. Arrows indicate increasing magnetic charge, and the gray underlay marks the type I-only part of the positive-$\zeta$ branch. The curves end at $Q_m/M=0.64,1.4,0.824$ for $\zeta/M^2=1,0,-1$, respectively, without including the limiting charges.}\label{fig:complexcomparison}
\end{figure}

Finally, we compare the complex-frequency trajectories in Fig.~\ref{fig:complexcomparison} with  the three backgrounds at $\alpha/M^2=5$. Each set of three curves starts from the same Schwarzschild frequency. In Fig.~\ref{fig:complexcomparison:a}, the negative-$\zeta$ dipole moves toward the real axis throughout the scan, including after its real frequency turns upward. The positive-$\zeta$ dipole first moves farther into the lower half-plane and then turns toward the real axis. On the GHS background, the dipole damping rate first increases slightly, then decreases, and finally rises sharply at large charge. The three backgrounds therefore give different damping trends after the real frequency passes through its minimum.
The two quadrupole branches are displayed  in Figs.~\ref{fig:complexcomparison:b} and \ref{fig:complexcomparison:c}. The G-led real frequencies decrease after a small initial increase. Their damping rates first increase and then decrease, with the largest variation on the GHS background. For the EM-led branch, the GHS real frequency increases to much larger values, while the positive-$\zeta$ real frequency varies over a narrow range. Near the last positive-$\zeta$ points, the dipole and gravitational-led quadrupole move toward the real axis, whereas the EM-led quadrupole moves away. Their decay times therefore change in opposite directions on the same background.

Varying $\zeta$ changes both the geometry and the electromagnetic coefficients. Comparisons at fixed charge must use the overlapping charge ranges, since the plotted endpoints correspond to different charges. The gray underlay marks the positive-$\zeta$ points with $Q_m/M=0.45$--$0.64$, where only the type I-local conditions hold. The curves stop below the electromagnetic bound for $\zeta>0$, the singular GHS endpoint for $\zeta=0$, and horizon merger for $\zeta<0$. These trajectories describe the fundamental branches continued from Schwarzschild at nonextremal charges. We do not extrapolate them to the endpoints or calculate additional mode families near horizon merger.

Consequently, we find  all negative imaginary frequencies for $\ell$=1,2(G,EM) and $\zeta/M^2=-1,1$, implying that the dEH black holes are stable against the type I perturbations. Simultaneously, we check that the magnetic  GHS black holes  with $\zeta=0$ is stable against the same perturbations.  

\FloatBarrier
\subsection{Numerical accuracy}\label{sec:verification}
We obtain every reported frequency independently with DI and PS. Tables~\ref{tab:frequencies} and \ref{tab:l3frequencies} compare representative $\ell=1,2$ and $\ell=3$ modes, respectively, at moderate charge and near the upper charge bounds. Both methods agree to the six decimal places displayed. The difference between the converged frequencies is
\begin{equation}
\epsilon_{\rm PD}=|M\omega_{\rm PS}-M\omega_{\rm DI}|,
\label{eq:methoddifference}
\end{equation}
evaluated before rounding. In Table~\ref{tab:frequencies}, the largest dipole and quadrupole differences are $2.75\times10^{-10}$ and $1.83\times10^{-9}$, respectively. The quadrupole maximum occurs for the G-led GHS mode at $Q_m/M=1.4$ and $\alpha=0$. The largest octupole difference in Table~\ref{tab:l3frequencies} is $1.30\times10^{-9}$, comparable to the quadrupole discrepancy. Near the type I bound at $Q_m/M=0.64$ and $\alpha/M^2=5$, the G-led and EM-led octupole differences are $4.07\times10^{-10}$ and $2.53\times10^{-10}$, respectively. The superscript $\dagger$ marks positive-$\zeta$ points satisfying only the type I local conditions, as shown in Fig.~\ref{fig:domain}.

\begin{table}[H]
\caption{Comparison between  DI and PS QNM frequencies for $\ell=1,2$, with $M=1$, $\alpha/M^2=0,1,5$, and $\zeta/M^2=0,1,-1$. The last column states the method difference defined in Eq.~\eqref{eq:methoddifference}.}\label{tab:frequencies}
\centering
\small
\setlength{\tabcolsep}{4.5pt}
\renewcommand{\arraystretch}{1.08}
{
\begin{tabular}{cccccc}\hline\hline
$\zeta/M^2$ & $\alpha/M^2$ & $Q_m/M$ & $M\omega_{\rm DI}$ & $M\omega_{\rm PS}$ & $\epsilon_{\rm PD}$\\\hline
\multicolumn{6}{c}{$\ell=1$, electromagnetic-led}\\\hline
$0$ & $0$ & $0.6$ & $0.279706-0.096392\ii$ & $0.279706-0.096392\ii$ & $2.01\times10^{-11}$\\
$0$ & $5$ & $1.4$ & $0.308043-0.127505\ii$ & $0.308043-0.127505\ii$ & $1.68\times10^{-10}$\\
$1$ & $1$ & $0.4$ & $0.257323-0.102320\ii$ & $0.257323-0.102320\ii$ & $1.20\times10^{-11}$\\
$1$ & $5$ & $0.64^{\dagger}$ & $0.286907-0.059718\ii$ & $0.286907-0.059718\ii$ & $2.75\times10^{-10}$\\
$-1$ & $5$ & $0.824$ & $0.258387-0.050433\ii$ & $0.258387-0.050433\ii$ & $2.11\times10^{-11}$\\
\hline
\multicolumn{6}{c}{$\ell=2$, gravitational-led}\\\hline
$0$ & $0$ & $1.4$ & $0.703216-0.084648\ii$ & $0.703216-0.084648\ii$ & $1.83\times10^{-9}$\\
$0$ & $5$ & $1.4$ & $0.289074-0.048037\ii$ & $0.289074-0.048037\ii$ & $1.56\times10^{-9}$\\
$1$ & $1$ & $0.4$ & $0.378409-0.089868\ii$ & $0.378409-0.089868\ii$ & $4.64\times10^{-12}$\\
$1$ & $5$ & $0.64^{\dagger}$ & $0.345556-0.070248\ii$ & $0.345556-0.070248\ii$ & $7.15\times10^{-11}$\\
$-1$ & $5$ & $0.824$ & $0.335645-0.084231\ii$ & $0.335645-0.084231\ii$ & $4.44\times10^{-10}$\\
\hline
\multicolumn{6}{c}{$\ell=2$, electromagnetic-led}\\\hline
$0$ & $0$ & $1.4$ & $1.251989-0.095120\ii$ & $1.251989-0.095120\ii$ & $6.38\times10^{-11}$\\
$0$ & $5$ & $1.4$ & $0.996868-0.095367\ii$ & $0.996868-0.095367\ii$ & $8.74\times10^{-10}$\\
$1$ & $1$ & $0.4$ & $0.470990-0.100549\ii$ & $0.470990-0.100549\ii$ & $1.84\times10^{-12}$\\
$1$ & $5$ & $0.64^{\dagger}$ & $0.444718-0.110421\ii$ & $0.444718-0.110421\ii$ & $3.46\times10^{-10}$\\
$-1$ & $5$ & $0.824$ & $0.484161-0.072751\ii$ & $0.484161-0.072751\ii$ & $5.80\times10^{-11}$\\
\hline
\hline\end{tabular}
}
\end{table}

\begin{table}[H]
\caption{Comparison between DI and PS QNM frequencies for the two $\ell=3$ branches, with $M=1$, $\alpha/M^2=0,1,5$, and $\zeta/M^2=0,1,-1$. The last column indicates  the difference between  two methods.}\label{tab:l3frequencies}
\centering
\small
\setlength{\tabcolsep}{4.5pt}
\renewcommand{\arraystretch}{1.08}
{
\begin{tabular}{cccccc}\hline\hline
$\zeta/M^2$ & $\alpha/M^2$ & $Q_m/M$ & $M\omega_{\rm DI}$ & $M\omega_{\rm PS}$ & $\epsilon_{\rm PD}$\\\hline
\multicolumn{6}{c}{$\ell=3$, gravitational-led}\\\hline
$0$ & $1$ & $1.4$ & $0.570017-0.133599\ii$ & $0.570017-0.133599\ii$ & $9.95\times10^{-11}$\\
$0$ & $5$ & $1.4$ & $0.456471-0.109674\ii$ & $0.456471-0.109674\ii$ & $1.03\times10^{-9}$\\
$1$ & $0$ & $0.4$ & $0.607914-0.093245\ii$ & $0.607914-0.093245\ii$ & $1.72\times10^{-11}$\\
$1$ & $5$ & $0.64^{\dagger}$ & $0.561419-0.101622\ii$ & $0.561419-0.101622\ii$ & $4.07\times10^{-10}$\\
$-1$ & $5$ & $0.824$ & $0.512653-0.091934\ii$ & $0.512653-0.091934\ii$ & $8.77\times10^{-10}$\\
\hline
\multicolumn{6}{c}{$\ell=3$, electromagnetic-led}\\\hline
$0$ & $1$ & $1.4$ & $1.423999-0.091932\ii$ & $1.423999-0.091932\ii$ & $1.30\times10^{-9}$\\
$0$ & $5$ & $1.4$ & $1.424042-0.091955\ii$ & $1.424042-0.091955\ii$ & $1.10\times10^{-10}$\\
$1$ & $0$ & $0.4$ & $0.688551-0.098344\ii$ & $0.688551-0.098344\ii$ & $4.56\times10^{-12}$\\
$1$ & $5$ & $0.64^{\dagger}$ & $0.662909-0.097186\ii$ & $0.662909-0.097186\ii$ & $2.53\times10^{-10}$\\
$-1$ & $5$ & $0.824$ & $0.720572-0.084344\ii$ & $0.720572-0.084344\ii$ & $2.07\times10^{-10}$\\
\hline
\hline\end{tabular}
}
\end{table}

\begin{figure}[htp]
\centering
\subfigure[\enspace $\ell=1$]{
\includegraphics[width=.32\textwidth]{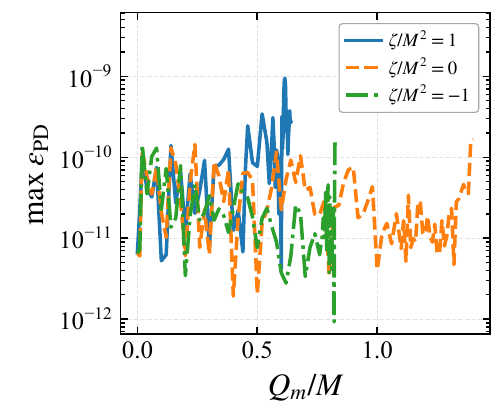}\label{fig:methoddifferences:a}}
\hfill
\subfigure[\enspace $\ell=2,3$, gravitational-led]{
\includegraphics[width=.32\textwidth]{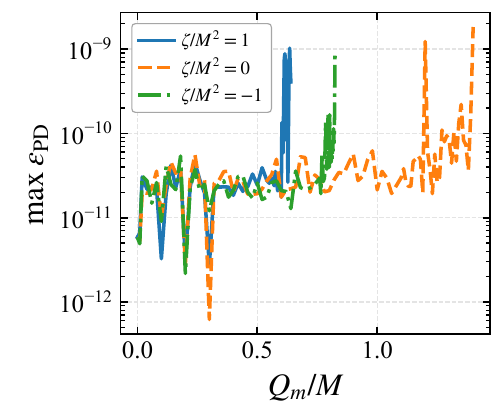}\label{fig:methoddifferences:b}}
\hfill
\subfigure[\enspace $\ell=2,3$, electromagnetic-led]{
\includegraphics[width=.32\textwidth]{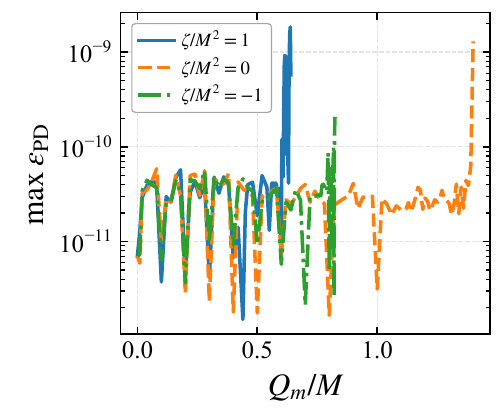}\label{fig:methoddifferences:c}}
\caption{Maximum DI--PS frequency differences over $\alpha/M^2=0,1,5$ for $M=1$ and $\zeta/M^2=1,0,-1$. Figs.~\ref{fig:methoddifferences:b} and \ref{fig:methoddifferences:c} also take the maximum over $\ell=2,3$. Solid, dashed, and dash-dotted curves denote $\zeta/M^2=1,0,-1$, respectively.}\label{fig:methoddifferences}
\end{figure}

\begin{figure}[htp]
\centering
\subfigure[\enspace PS refinement]{
\includegraphics[width=.485\textwidth]{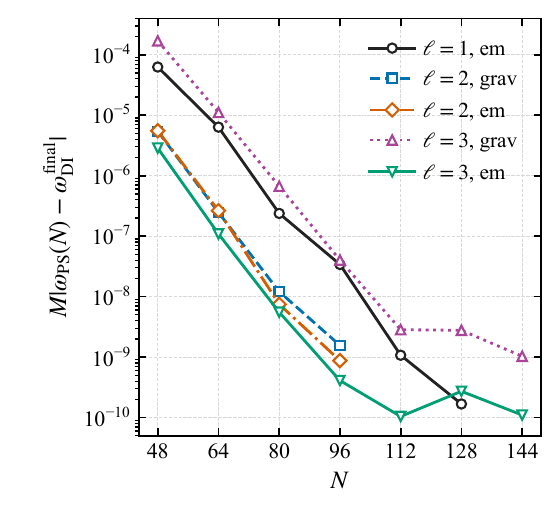}\label{fig:hardconvergence:a}}
\hfill
\subfigure[\enspace DI refinement]{
\includegraphics[width=.485\textwidth]{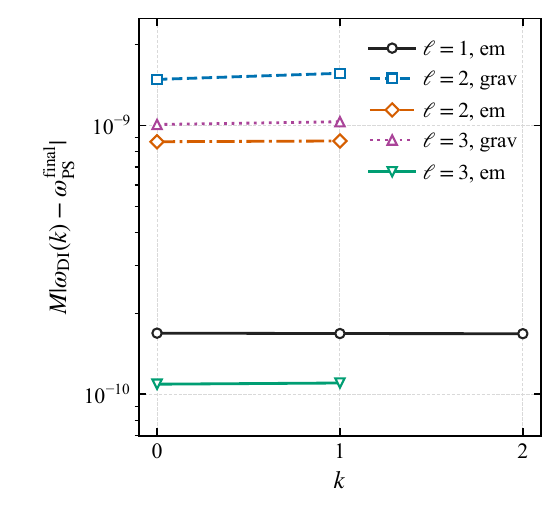}\label{fig:hardconvergence:b}}
\caption{Convergence of the five type I branches for $M=1$, $\zeta=0$, $\alpha=\beta=5$, and $Q_m/M=1.4$. Figs.~\ref{fig:hardconvergence:a} and \ref{fig:hardconvergence:b} show differences from the final frequency of the other method as the PS node number $N$ and DI refinement index $k$ increase, respectively. { The labels ``grav'' and ``em'' denote the G-led and EM-led branches.}}\label{fig:hardconvergence}
\end{figure}

Fig.~\ref{fig:methoddifferences} summarizes the agreement over the full charge scans, where the largest DI--PS difference amounts to  $1.842\times10^{-9}$. The curves give the maximum difference over the couplings, and the two coupled-mode panels also take the maximum over $\ell=2,3$. Changes between the last two PS and DI refinements are below $1.331\times10^{-8}$ and $1.059\times10^{-9}$, respectively. We evaluate the PS radial-equation residual at interior test points and normalize it by the summed norms of the differential-equation terms. This residual and the DI matching residual are both below $10^{-8}$. These checks support the reported six-decimal precision, including near the charge bounds. Figure~\ref{fig:hardconvergence} illustrates convergence for all five branches on the GHS background at $\alpha=\beta=5M^2$ and $Q_m/M=1.4$. The two panels vary the PS node number $N$ and the DI refinement index $k$, respectively. For these five modes, the final DI--PS differences range from approximately $1.10\times10^{-10}$ to $1.56\times10^{-9}$. At large $N$, the PS curves approach a numerical plateau with small nonmonotonic changes. The DI curves are nearly horizontal because further DI refinement changes the frequencies much less than the remaining difference from PS. These comparisons provide numerical consistency checks rather than rigorous error bounds. Refinement effects are much smaller than the frequency shifts and damping variations discussed above. The numerical settings and additional checks are given in Appendix~\ref{sec:methods}.

\FloatBarrier
\clearpage

\section{Conclusions}\label{sec:conclusion}
The coupled QNM problem of magnetically charged dEH black holes has been studied in the type I sector, where axial gravitational perturbations couple to polar electromagnetic perturbations. The dipole sector reduces to a single electromagnetic equation, while the higher multipoles ($\ell\ge2$) contain two coupled radiative degrees of freedom. The fundamental branches continued from Schwarzschild have been computed for $\ell=1,2,3$ by direct integration and Chebyshev pseudospectral methods, which give consistent frequencies throughout the parameter ranges.

The magnetic background depends on $\zeta=\alpha-\beta$, whereas the polar electromagnetic perturbations also depend on $\alpha$ through $\Delta_2$. At finite charge, varying $\alpha$ at fixed $\zeta$, mass, and charge therefore shifts both coupled branches for $\ell\ge2$ on the same geometry, since each mode contains gravitational and electromagnetic components. Along the GHS charge sequence, the G-led quadrupole damping is nonmonotonic. At the largest charges on the GHS and positive-$\zeta$ backgrounds, increasing $\alpha/M^2$ from 0 to 5 produces opposite G-led damping shifts for $\ell=2$ and 3. Along the $\zeta/M^2=-1$ sequence, the dipole damping decreases as the charge approaches the horizon-merger limit. These results show that a decrease in the oscillation frequency does not necessarily imply a longer decay time. The ranges end below charge limits set by $\Delta_1(\rh)=0$ at a nonextremal horizon for $\zeta>0$, the singular GHS endpoint for $\zeta=0$, and horizon merger for $\zeta<0$. For $\zeta>0$, the type I charge range extends into a region where the complementary electromagnetic polarization violates its angular-gradient condition. All calculated fundamental branches remain damped throughout these nonextremal ranges, which supports stability against the type I modes considered here.

Further work may extend the calculation to higher multipoles and overtones, which would clarify how the coupling dependence varies across the type I spectrum. Time-domain evolution would provide a direct picture of the excitation and decay of the coupled gravitational and electromagnetic perturbations, allowing their contributions to the ringdown signal to be compared. A dedicated search for additional mode families near horizon merger would also help identify which modes dominate the ringdown as the black hole approaches extremality.

\FloatBarrier

\appendix
\section{Reduction of the perturbation equations}\label{app:derivation}

For the perturbations in Eqs.~\eqref{eq:metricpert} and \eqref{eq:EMpert}, $\delta X=0$ and $\overline Z=0$, but
\begin{equation}
\delta Z=\delta\!\left(F_{\mu\nu}\widetilde F^{\mu\nu}\right)
=\frac{4Q_m}{r^2}\sqrt{\frac{B}{A}}
\left(u_1'+\ii\omega u_2\right)Y_{\ell0}e^{-\ii\omega t},
\label{eq:dualvariation}
\end{equation}
where the sign follows from $\overline F_{\theta\varphi}=Q_m\sin\theta$, $\delta F_{tr}=-(u_1'+\ii\omega u_2)Y_{\ell0}e^{-\ii\omega t}$, and the dual convention in Sec.~\ref{sec:field-equations}. The linear variation of the $\alpha f(\phi)Z\widetilde F^{\mu\nu}$ term in Eq.~\eqref{eq:Maxwellfield} therefore survives and produces the $\alpha$ dependence through $\Delta_2$ in Eq.~\eqref{eq:D123}.

\subsection{Dipole reduction}\label{app:dipolereduction}
For $\ell=1$, the axial tensor harmonic vanishes,
$D_{(a}(\varepsilon_{b)}{}^cD_cY_{10})=0$,
where $a,b,c$ are angular indices, $\varepsilon_{ab}$ is the unit-sphere Levi--Civita tensor, and parentheses denote symmetrization. The corresponding angular Einstein equation is absent.

For $Q_m\ne0$ and $\omega\ne0$, the remaining constraints and Eq.~\eqref{eq:Gmaster} give
\begin{align}
u_1'+\ii\omega u_2&=\frac{e^\phi}{2Q_m\Delta_2}
\sqrt{\frac{A\Delta_1}{B}}\Psi_{\rm em},
\label{eq:dipoleconstraint}\\
u_2&=-\frac{Q_mh_1}{r^2}
-\frac{\ii\omega e^\phi r^2}{4Q_m\sqrt{AB\Delta_1}}\Psi_{\rm em},
\label{eq:dipoleu2}\\
u_1&=-\frac{Q_mh_0}{r^2}
+\frac{e^\phi r^2}{4Q_m}\sqrt{\frac{AB}{\Delta_1}}
\left[\Psi_{\rm em}'
+\left(-\phi'-\frac2r+\frac{\Delta_1'}{2\Delta_1}\right)\Psi_{\rm em}\right].
\label{eq:dipoleu1}
\end{align}
Substituting Eqs.~\eqref{eq:dipoleu1} and \eqref{eq:dipoleu2} into Eq.~\eqref{eq:dipoleconstraint} and eliminating $h_0'+\ii\omega h_1-2h_0/r$ with Eq.~\eqref{eq:Gmaster} gives Eq.~\eqref{eq:dipolemaster} and the potential~\eqref{eq:dipolepotential}. The reconstruction requires $Q_m\ne0$, while the master equation has the smooth Schwarzschild limit~\eqref{eq:dipoleneutrallimit}.

\subsection{Higher-multipole reduction}\label{app:higherreduction}
For $\ell\ge2$ and $\omega\ne0$, the independent radial Einstein and Maxwell equations form the first-order system
\begin{align}
h_0'={}&\frac{2h_0}{r}
+\frac{\ii}{\omega}\left[\frac{A[\ell(\ell+1)-2]}{r^2}
+\frac{4AQ_m^2e^{-2\phi}\Delta_1}{r^8}-\omega^2\right]h_1\nonumber\\
&+\frac{4\ii AQ_me^{-2\phi}\Delta_1}{\omega r^6}u_2,
\label{eq:FOh0}\\
h_1'={}&-\frac{(AB)'}{2AB}h_1-\frac{\ii\omega}{AB}h_0,
\label{eq:FOh1}\\
u_1'={}&-\ii\omega u_2
+\frac{\ii\ell(\ell+1)A\Delta_1}{\omega r^4\Delta_2}
\left(Q_mh_1+r^2u_2\right),
\label{eq:FOu1}\\
u_2'={}&-\frac{\ii\omega u_1}{AB}-\frac{(AB)'u_2}{2AB}\nonumber\\
&+\left(\frac4r+2\phi'-\frac{\Delta_1'}{\Delta_1}\right)
\left(u_2+\frac{Q_mh_1}{r^2}\right)+\frac{2Q_mh_1}{r^3}.
\label{eq:FOu2}
\end{align}
The other linearized components follow from these relations and the background field equations. Solving Eqs.~\eqref{eq:FOh1} and \eqref{eq:FOu2} for $h_0$ and $u_1$ gives
\begin{align}
h_0&=\frac{\ii}{\omega}\left[ABh_1'+\frac{(AB)'}2h_1\right],\label{eq:reconstructh0}\\
u_1&=\frac{\ii AB}{\omega}\Bigg[u_2'+\frac{(AB)'u_2}{2AB}\nonumber\\
&\qquad-\left(\frac4r+2\phi'-\frac{\Delta_1'}{\Delta_1}\right)
\left(u_2+\frac{Q_mh_1}{r^2}\right)-\frac{2Q_mh_1}{r^3}\Bigg].
\label{eq:reconstructu1}
\end{align}
Substitution into Eqs.~\eqref{eq:FOh0} and \eqref{eq:FOu1} yields Eqs.~\eqref{eq:h1radial} and \eqref{eq:u2radial}, respectively. At $Q_m=0$, the coupling terms vanish and the equations reduce to the Schwarzschild gravitational and electromagnetic equations with the potentials in Eq.~\eqref{eq:Schwpotentials}.

\section{Numerical methods}\label{sec:methods}

Both methods use $\bm y=\Psi_{\rm em}$ for $\ell=1$ and $\bm y=(Ah_1/r,Au_2)^{\rm T}$ for $\ell\ge2$, with the boundary conditions~\eqref{eq:QNMbc}. All calculations use double-precision arithmetic. Branches are followed using frequency continuity and normalized eigenfunction overlap, with charge steps no larger than $0.025M$ for the initial continuation and $0.02M$ for the final dense scans. The label $n=0$ refers to the fundamental Schwarzschild branches, which need not remain the least damped modes at finite charge.

\subsection{Direct integration (DI)}
The DI method matches independent ingoing and outgoing solutions~\cite{Chandrasekhar:1975zza,Pani:2013pma}. For a nonextremal horizon with surface gravity $\kappa=\sqrt{A'(\rh)B'(\rh)}/2$, the expansions implementing Eq.~\eqref{eq:QNMbc} are
\begin{align}
\bm y_{\rm in}&=\left(\frac{r-\rh}{\rh}\right)^{-\ii\omega/(2\kappa)}
\sum_{j=0}^{p}\bm a_j\left(\frac{r-\rh}{\rh}\right)^j,\label{eq:DIhorizon}\\
\bm y_{\rm out}&=e^{\ii\omega r}\left(\frac rM\right)^{2\ii M\omega}
\sum_{j=0}^{p}\bm b_j\left(\frac{\rh}{r}\right)^j.
\label{eq:DIinfinity}
\end{align}
The coefficients follow recursively from the radial equations, and $p$ is the truncation order. For the dipole, all coefficients are scalars. The ingoing solutions are integrated along the real exterior. The outgoing solutions are integrated from $r_0=r_{\rm match}+L_{\rm out}e^{\ii\theta}$ back to the real matching radius $r_{\rm match}$ along a straight complex contour of length $L_{\rm out}$. Choosing $\operatorname{Im}(\omega e^{\ii\theta})>0$ makes the outgoing exponential decay toward the outer endpoint. The background square root is continued from the real exterior on the branch $\sqrt{Q_m^4+4M^2r^2}/(2Mr)\to1$, avoiding coefficient singularities.

The columns of $\bm Y_{\rm in}$ and $\bm Y_{\rm out}$ are independent solutions. A QNM occurs when their solution spaces intersect at $r_{\rm match}$:
\begin{equation}
\det\!
\begin{pmatrix}\bm Y_{\rm in}&\bm Y_{\rm out}\\
\bm Y_{\rm in}'&\bm Y_{\rm out}'\end{pmatrix}_{r=r_{\rm match}}=0.
\label{eq:DImatching}
\end{equation}
The matching matrix is $2\times2$ for $\ell=1$ and $4\times4$ for $\ell\ge2$. Orthonormal bases are used when necessary. DI refinement increases $p$ from 12 to 16 and extends $L_{\rm out}$ from $110M$ to $140M$ or farther, while varying $r_{\rm match}$ and $\theta$. Near a charge bound, the horizon starting offset and series radius are reduced to stay within the nearest coefficient singularity. In Fig.~\ref{fig:hardconvergence:b}, $k=0,1$ correspond to $p=12,14$ for all branches; the dipole also has $k=2$ with $p=16$.

\subsection{Pseudospectral (PS) method}
The exterior is compactified with $x=(r-\rh)/(r-\rh+L)\in[0,1]$, where $L>0$. Factoring out the QNM asymptotics gives~\cite{Boyd2001,Trefethen2000,grandclment2009spectral}
\begin{equation}
\bm y(r)=e^{\ii\omega r}(1-x)^{-2\ii M\omega}x^{-\ii\omega/(2\kappa)}\bm v(x).
\label{eq:PSfactor}
\end{equation}
The prefactor has the ingoing and outgoing behavior required by Eq.~\eqref{eq:QNMbc} at $x=0$ and 1, respectively. The regular function $\bm v$ is approximated on $N$ open Chebyshev--Gauss nodes $x_j=[1+\cos((2j+1)\pi/(2N))]/2$, with $j=0,\ldots,N-1$. Barycentric differentiation gives the quadratic eigenvalue problem
\begin{equation}
\left(\bm M_0+\omega\bm M_1+\omega^2\bm M_2\right)\bm v_N=0,
\label{eq:pencil}
\end{equation}
where $\bm v_N$ collects the nodal field values. The matrices are independent of $\omega$, and the quadratic problem is solved as a generalized eigenvalue problem~\cite{TisseurMeerbergen2001}.

PS refinement increases $N$ from 48 to 96, with further increases up to 160 where needed. In Fig.~\ref{fig:hardconvergence:a}, $N$ increases from 48 in steps of 16 to final values $128,96,144$ for $\ell=1,2,3$, respectively. The radial-equation residual is evaluated at interior points between nodes and normalized by the summed norms of the differential-equation terms. Near the charge bounds, the compactification length $L$ is also varied. Only frequencies that converge under refinement and agree with DI are retained, as discussed in Sec.~\ref{sec:verification}.

For the tested $\ell=3$ modes near the charge bounds, further DI boundary refinements change $M\omega$ by less than $2.53\times10^{-10}$, while varying $L$ in the PS method changes it by less than $1.33\times10^{-9}$. These additional checks support the numerical comparisons in Sec.~\ref{sec:verification}.

\clearpage
\begingroup
\linespread{1.3}\selectfont
\bibliographystyle{unsrtnat}
\bibliography{axial-polar-QNM}
\endgroup
\end{document}